\documentclass[twocolumn,preprintnumbers,superscriptaddress,prb]{revtex4}
\usepackage{graphicx,bm,times,url}
\usepackage{amsmath}
\usepackage{amsfonts}
\usepackage{epstopdf}
\usepackage{natbib}
\usepackage{verbatim}
\graphicspath{{./fig/}{./png/}}

\newcommand{\BB}{\bm{B}}
\newcommand{\JJ}{\mbox{\boldmath $J$} {}}
\newcommand{\uu}{\mbox{\boldmath $u$} {}}
\newcommand{\nab}{\mbox{\boldmath $\nabla$} {}}
\newcommand{\AAA}{\mbox{\boldmath $A$} {}}
\newcommand{\EQ}{\begin{equation}}
\newcommand{\EN}{\end{equation}}
\newcommand{\ee}{\mbox{\boldmath $e$} {}}
\newcommand{\dd}{{\rm d} {}}
\newcommand{\rr}{\mbox{\boldmath $r$} {}}
\newcommand{\sss}{\mbox{\boldmath $s$} {}}
\def\Tr{\mbox{\rm Tr}}
\def\cs{c_{\rm s}}
\newcommand{\Fig}[1]{Figure~\ref{#1}}
\newcommand{\xx}{\bm{x}}
\newcommand{\FF}{\mbox{\boldmath $F$} {}}
\newcommand{\yy}{\mbox{\boldmath $y$} {}}

\newcommand{\refbold}[1]{#1}
\begin{document}

\title{Ideal Relaxation of the Hopf Fibration} 

\author{Christopher Berg Smiet}
\affiliation{Huygens-Kamerlingh Onnes Laboratory, Leiden University, P.O.\ Box 9504, 2300 RA Leiden, The Netherlands}
\author{Simon Candelaresi}
\affiliation{Division of Mathematics, University of Dundee, Dundee DD1 4HN, United Kingdom}
\author{Dirk Bouwmeester}
\affiliation{Huygens-Kamerlingh Onnes Laboratory, Leiden University, P.O.\ Box 9504, 2300 RA Leiden, The Netherlands}
\affiliation{Department of Physics, University of California Santa Barbara, Santa Barbara, California, 93106, USA}

\date{\today}

\begin{abstract}
    \refbold{Ideal MHD relaxation is the topology-conserving reconfiguration of a magnetic
    field into a lower energy state where the net force is zero. This is achieved by modeling
    the plasma as perfectly conducting viscous fluid.
    It is} an important tool for investigating plasma equilibria and
    is often used to study the magnetic configurations in fusion devices and astrophysical
    plasmas.
    We study the equilibrium reached by a localized
    magnetic field through the topology conserving relaxation of a magnetic field based
    on the Hopf fibration in which magnetic field lines
    are closed circles that are all linked with one another.
    Magnetic fields with this topology have recently been shown to occur in
    non-ideal numerical simulations.
    Our results show that any localized field can only attain equilibrium if there is a
    finite external pressure, and that for such a field
    a Taylor state is unattainable.
    We find an equilibrium plasma configuration that is characterized by a lowered
    pressure in a toroidal region, with field lines lying on surfaces of
    constant pressure.
    Therefore, the field is in a Grad-Shafranov equilibrium.
    Localized helical magnetic fields are found when
    plasma is ejected from astrophysical bodies and subsequently relaxes
    against the background plasma, as well as on earth in plasmoids generated by e.g.\ a Marshall
    gun.
    This work shows under which conditions an equilibrium can be reached and identifies a
    toroidal depression as the characteristic feature of such a configuration.
\end{abstract}

\maketitle

\section{Introduction}
A fundamental question in plasma physics is: Given a magnetic field configuration,
\refbold{what equilibrium state can it attain?}
This question was posed by
Arnol'd \cite{arnold1974asymptotic} who considered the \refbold{static equilibria} of ideal
(zero magnetic \refbold{diffusivity}),
incompressible magnetohydrodynamics (MHD)\refbold{, such that the magnetic topology remains
unchanged}.
Subsequent work by Moffatt expanded this problem for various geometries and
connected the equilibrium solutions to solutions of Euler equations
for fluid flow \citep{moffatt1985magnetostatic}.
J.B.\ Taylor \citep{taylor1974relaxation} considered the problem for a different scenario;
a plasma with a very low (but finite) resistivity in a toroidal device.
His conjecture was both elegant and experimentally accurate: the field relaxes to a linear
force-free state with the same helicity of the initial field. \refbold{Taylor's theory is  an application
of the work of Woltjer, who showed that a force-free (Beltrami) state is the lowest energy
configuration that a field can attain under conservation of
helicity\cite{woltjer1958theorem, woltjer1958hydromagnetic, woltjer1959hydromagnetic}.}

Due to the elegance and predictive power of Taylor's conjecture, this principle of relaxation
to a linear force-free state is often applied also in geometries beyond which it is strictly
applicable.
Recently there have been several papers addressing this, and identifying geometries in which
the final state after relaxation is distinctly not a Taylor state.
Simulations on magnetic field relaxation in a flux tube geometry have shown
additional topological constraints
associated with the field line connectivity that hinder relaxation to a force-free state
\cite{Yeates_Topology_2010}.
Also in one-dimensional resistive simulations fields were found not to converge to a linear
force-free
state \cite{moffatt2015magnetic}. Furthermore, in our recent work we investigated the resistive
decay of linked flux rings and tubes that converge to an MHD equilibrium that is not
force-free \cite{Smiet-Candelaresi-2015-115-5-PRL}.

The magnetic topology of this last example is remarkable, the field is localized,
has finite energy, and
field lines lie on nested toroidal surfaces such that on each surface the ratio of poloidal to
toroidal winding is nearly identical.
This last observation implies that the magnetic field topology is related to the
mathematical structure called the Hopf
map\citep{Hopf-1931-MatAn}.
Fibers of this map form circles that are all linked with each other, and lie on
nested toroidal surfaces.
The structure of the Hopf map has been used in many branches of physics, amongst others,
to describe structures in
superfluids \cite{volovik1977} and spinor Bose-Einstein condensates \citep{Kawaguchi2008, hall2016tying}.
It also forms the basis for new analytical solutions to Maxwell's equations \cite{Ranada1989, Irvine2008}
and Einstein's equations \cite{thompson2015classification}.
In ideal, incompressible MHD the Hopf map has been used to generate solutions of
the ideal MHD equations called topological solitons \citep{kamchatnov1982topological,
Sagdeev1986nonlinear}.

Even though the localized MHD equilibrium\cite{Smiet-Candelaresi-2015-115-5-PRL} has similar
magnetic topology to the Hopf fibration, the geometry is different.
The equilibrium consists of a balance between the pressure gradient force, directed inwards
towards the magnetic axis, and the Lorentz force, directed outwards.
In this paper we investigate exactly this equilibrium, and how it geometrically relates to
fields derived from the Hopf map.
We take fields with these well-defined
magnetic topologies, and find their equilibrium configurations using
\refbold{a relaxation} method that exactly conserves field line topology \refbold{and converges to a
static equilibrium which is a solution of the ideal MHD equations}.

The choice for this initial topology is inspired by the numerical
results on linked rings\cite{Smiet-Candelaresi-2015-115-5-PRL}, but there are many
other works in which localized MHD equilibria are investigated and to which our results apply.
Localized magnetic structures have been described in numerical relaxation experiments
and are referred to as magnetic bubbles \citep{gruzinov2010solitary,
braithwaite2010magnetohydrodynamic, zrake2016freely}.
Also in fusion research structures are described as compact toroids or plasmoids,
which consist of magnetic field lines lying on closed surfaces\cite{bostick1956experimental}.
Sometimes these structures are described as embedded in a guide field, but
in isolation these fields are localized and show a similar magnetic topology
\citep{armstrong1980compact,  perkins1988deep, wright1990field}.
Some models for magnetic clouds, regions of increased magnetic field observed in the solar
wind\cite{burlaga1991magnetic}, consider the cloud as a localized magnetic excitation, either a
current-ring \cite{kumar1996interplanetary}, or a flare-generated
spheromak\cite{ivanov1985interplanetary}.

In ideal MHD magnetic helicity, or linking of magnetic field lines, is exactly conserved.
Magnetic helicity is defined as
\EQ
H_{\rm M}=\int \AAA\cdot \BB \ \mathrm{d}^3x,
\EN
where $\AAA$ is the vector potential and $\BB = \nab\times\AAA$ the magnetic field.
Woltjer was the first to realize that the value of this integral
is conserved in ideal MHD \citep{woltjer1958theorem}.
It has recently been shown that \refbold{any regular integral invariant under volume-preserving
transformations is equivalent to the helicity}\citep{enciso2016helicity}.
Moffatt \citep{moffatt1969degree, arnold1974asymptotic} gave
helicity a topological interpretation; helicity is a measure for
the self- and inter-linking of magnetic field lines in a plasma.
The conservation of magnetic linking can also be physically understood by the fact that
in a perfectly conducting fluid the magnetic flux through a fluid element cannot change, and
the magnetic field is transported by the fluid flow, a condition
referred to as the frozen in condition \citep{alfven1942existence, batchelor1950spontaneous,
PriestReconnection2000}.
As a consequence, in ideal MHD any linking or knottedness of  magnetic field lines
cannot be undone, and the magnetic topology\cite{hornig1996magnetic}
is conserved.
Ideal MHD thus conserves not only total magnetic helicity but also the linking of every field
line with every other field line.

Simulating non-resistive MHD numerically using a fixed Eulerian grid is a notoriously difficult problem
due to numerical \refbold{errors} in Eulerian finite difference schemes \cite{NumericalRecipes3}.
It is possible to circumvent this by using a Lagrangian relaxation scheme
\citep{Pontin-Hornig-2009-700-2-ApJ} \refbold{which dissipates fluid motion} but perfectly preserves the
frozen in condition.
This was recently implemented using
mimetic numerical operators in the numerical code GLEMuR\citep{Candelaresi-2014-mimetic}.
In this paper we study the non-resistive relaxation of magnetic fields with the topology of
the Hopf map using this recently developed code.
\refbold{Lagrangian methods were also recently implemented in a 2d dissipationless ideal MHD
evolution scheme to study current singularities \cite{zhou2014variational}.}

\refbold{The virial theorem of MHD is a useful tool to investigate possible MHD equilibrium
configurations}\cite{chandrasekhar1961hydrodynamic, kulsrud2005plasma}.
This theorem relates the second derivative of the moment of inertia $I$ to integrals
over the volume and boundary of a region in the plasma, and is usually stated as:
\EQ
\frac{\dd^2 I}{\dd t^2} = -\int\limits_{\partial V}\mathbf{T}\cdot\rr \cdot \dd\sss +
\int\limits_{V}\Tr(\mathbf{T})\ \dd^3x.
\label{eq: virial tensor}
\EN
Here $\Tr(\mathbf{T})$ denotes the trace of the strain tensor
$\mathbf{T}=\mathbf{T}_{\uu}+\mathbf{T}_p+\mathbf{T}_{\BB}$.
This tensor has a velocity
component $\mathbf{T}_{\uu}=\rho \uu \uu$, a pressure component $\mathbf{T}_p=\mathbf{I}p$
and a component due to the magnetic forces $\mathbf{T}_{\BB}=\mathbf{I}B^2/2- \BB \BB$. Here $\rho$
denotes the fluid density, $\uu$ the velocity and $V$ the domain.
$\rr$ is the position vector and $\sss$ indicates the surface normal of the surface of the region.

A consequence of the virial theorem is that for any static equilibrium to exist ($I$ to remain constant),
the contribution of the bulk must be compensated by corresponding stresses on the boundary.
The integral over the bulk can be written as $\int_V \left( \rho u^2+3p +B^2\right) \dd^3 x $,
which is always positive.
Any reorganization of the bulk can change the magnitude of this contribution, but it will always be finite,
which implies that without any stresses on the surface, a plasma will always expand ($I$ will increase).
\refbold{Therefore, one of the surface terms must integrate to a non-zero value for any equilibrium.}

If we consider a localized magnetic field, such as the Hopf field, which has the same magnetic
field topology as the localized equilibria
described in previous numerical experiments\cite{Smiet-Candelaresi-2015-115-5-PRL},
then their magnetic field strength vanishes at sufficient
distance, where we can put our boundary.
This leaves two possible configurations through which an equilibrium can be reached.
The first configuration has a finite pressure at the boundary.
Any expansion in the bulk will create a low-pressure
region, which will prevent the structure from expanding indefinitely.
The second configuration has finite magnetic stresses at the boundary.
This can be achieved by adding a
constant guide field that prevents the field from expanding indefinitely through magnetic tension
from the guide field.
We note that the first configuration can never converge to a Taylor state, i.e.\ a localized magnetic field
cannot relax to a force-free configuration.

\section{The Hopf Field} \label{sec: hopf field}
\begin{figure}[t!]\begin{center}
\includegraphics[width=0.9\columnwidth]{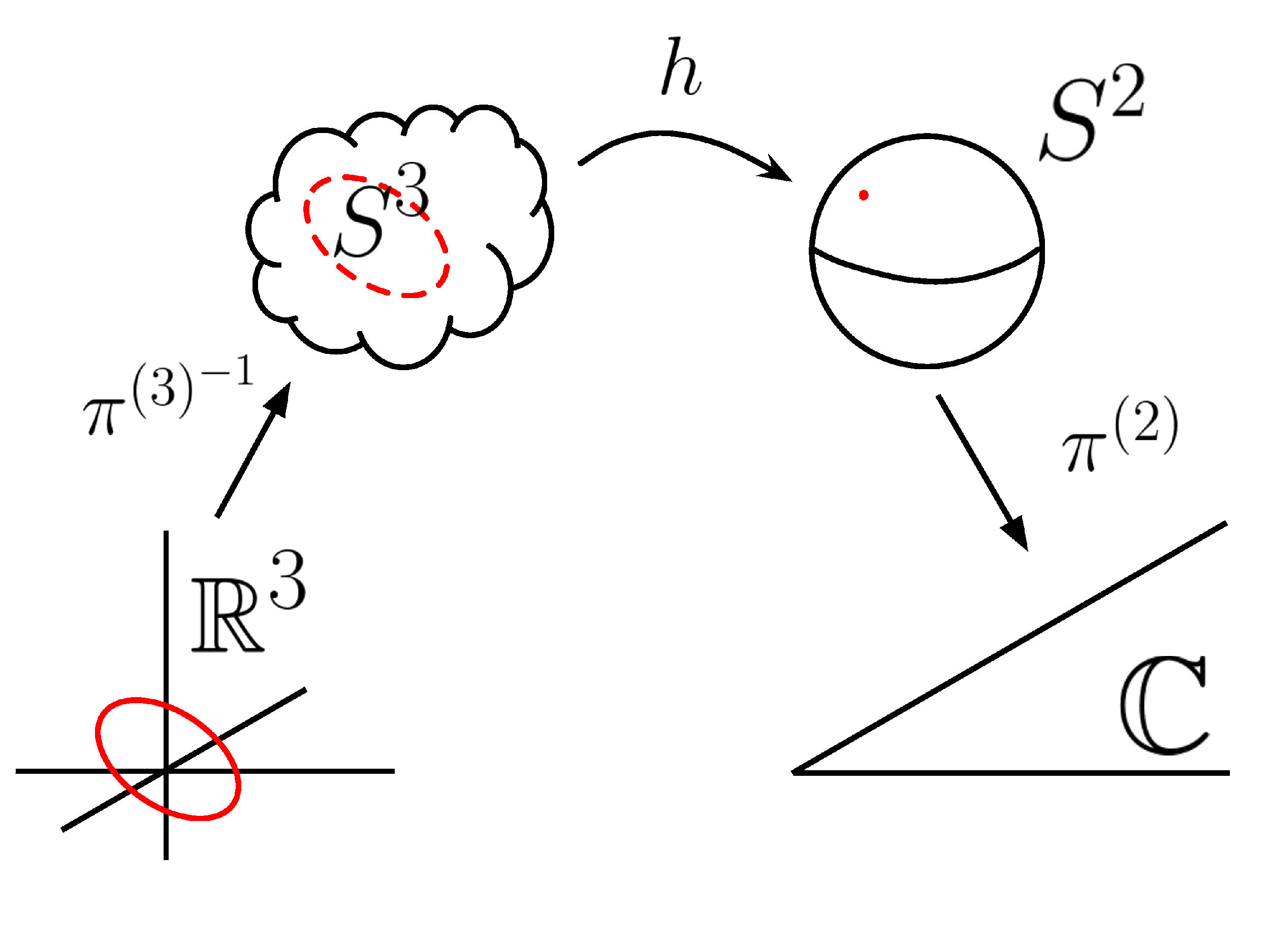}
\end{center}
\caption[]{
Illustration of the construction of a field with the topology of the Hopf map.
The red dashed circle in $S^3$ is a fiber of the map. Through stereographic projection the fiber
structure of the Hopf map is translated to $\mathbb{R}^3$ and a field that lies everywhere tangent to these
circles is constructed.
}
\label{fig: hopfmap}
\end{figure}

In 1931 Heinz Hopf \citep{Hopf-1931-MatAn}
discovered a curious property of maps
from the hypersphere $S^3$ \refbold{onto} the sphere $S^2$, namely that the fibers of the
maps (pre-images of points on $S^2$) are circles in $S^3$ that are all linked.
This class of functions can be extended to a function from $\mathbb{R}^3$ to $\mathbb{C}$
from which a divergence-free vector field in $\mathbb{R}^3$ can be constructed
such that the integral curves (field lines) lie tangent to the original fibers of the map
\citep{Sagdeev1986nonlinear, Ranada1989, kamchatnov1982topological}.
This construction is illustrated in \Fig{fig: hopfmap}.

The Hopf map can be modified as described in \cite{arrayas2014class},
such that every fiber of the map
lies on a toroidal surface with poloidal winding $\omega_1$ (short way around the torus)
and toroidal winding $\omega_2$ (long way around the torus).
If $\omega_1$ and $\omega_2$ are commensurable
($\omega_1/\omega_2 \in \mathbb{Q}$), all field lines are $(\omega_1/{\rm gcd}(\omega_1,\omega_2), \omega_2/{\rm gcd}(\omega_1, \omega_2))$ torus
knots where ${\rm gcd}(a, b)$ is the greatest common divisor of $a$ and $b$.
From this map a field in $\mathbb{R}^3$ can be generated with that magnetic topology.
Every field line lies on a torus and the tori form a nested set filling all of space.
In this field there are two special field lines that do not form a $(\omega_1,\omega_2)$ torus knot.
One lies on the largest torus, which reduces to a straight field line on the $z$-axis
(torus through infinity), and the other field line lies
on the degenerate (smallest) torus that reduces to a unit circle in the $xy$-plane and that is called
the degenerate field line.

\begin{figure}[t!]\begin{center}
\includegraphics[width=0.9\columnwidth]{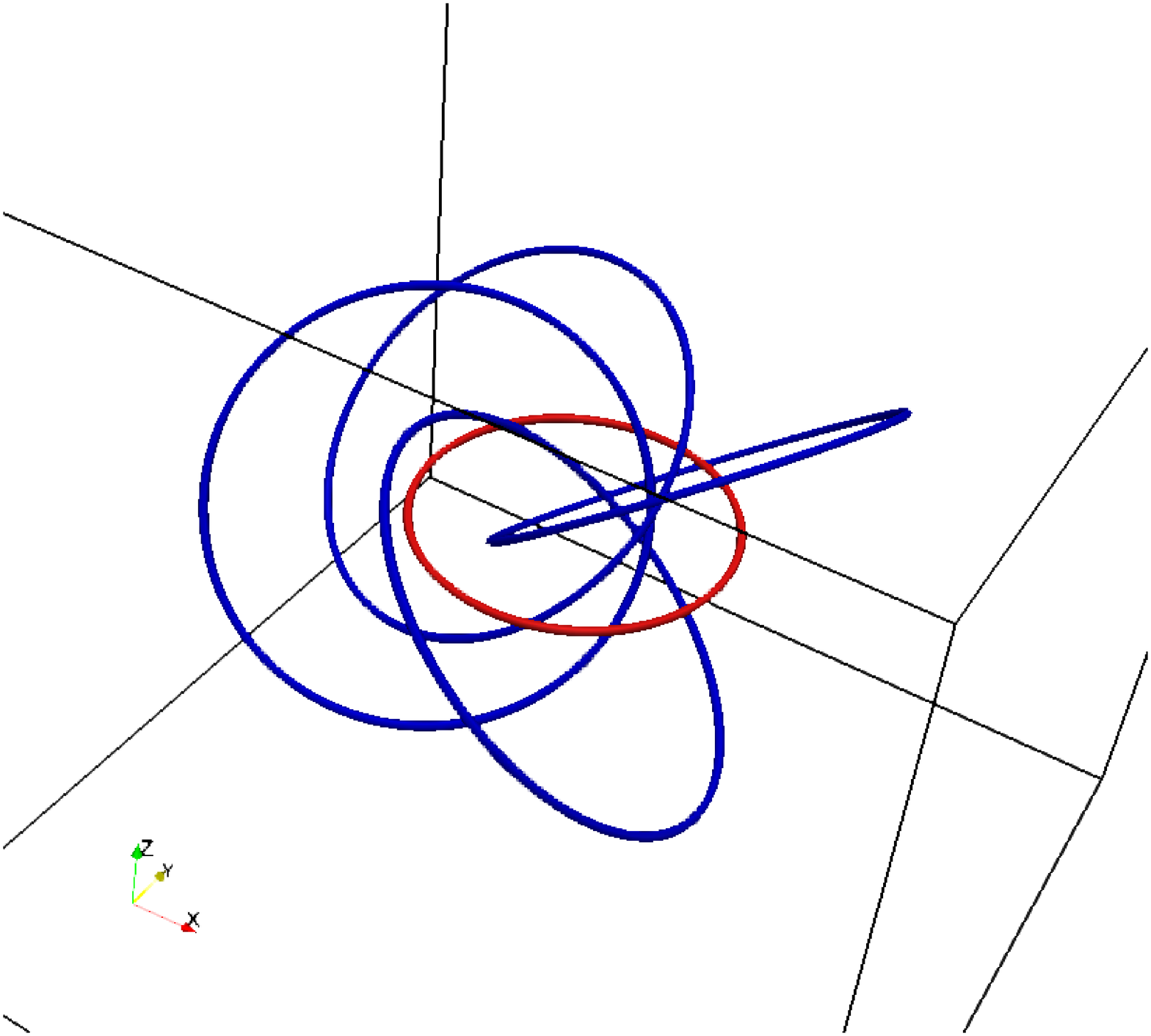} \\
\includegraphics[width=0.9\columnwidth]{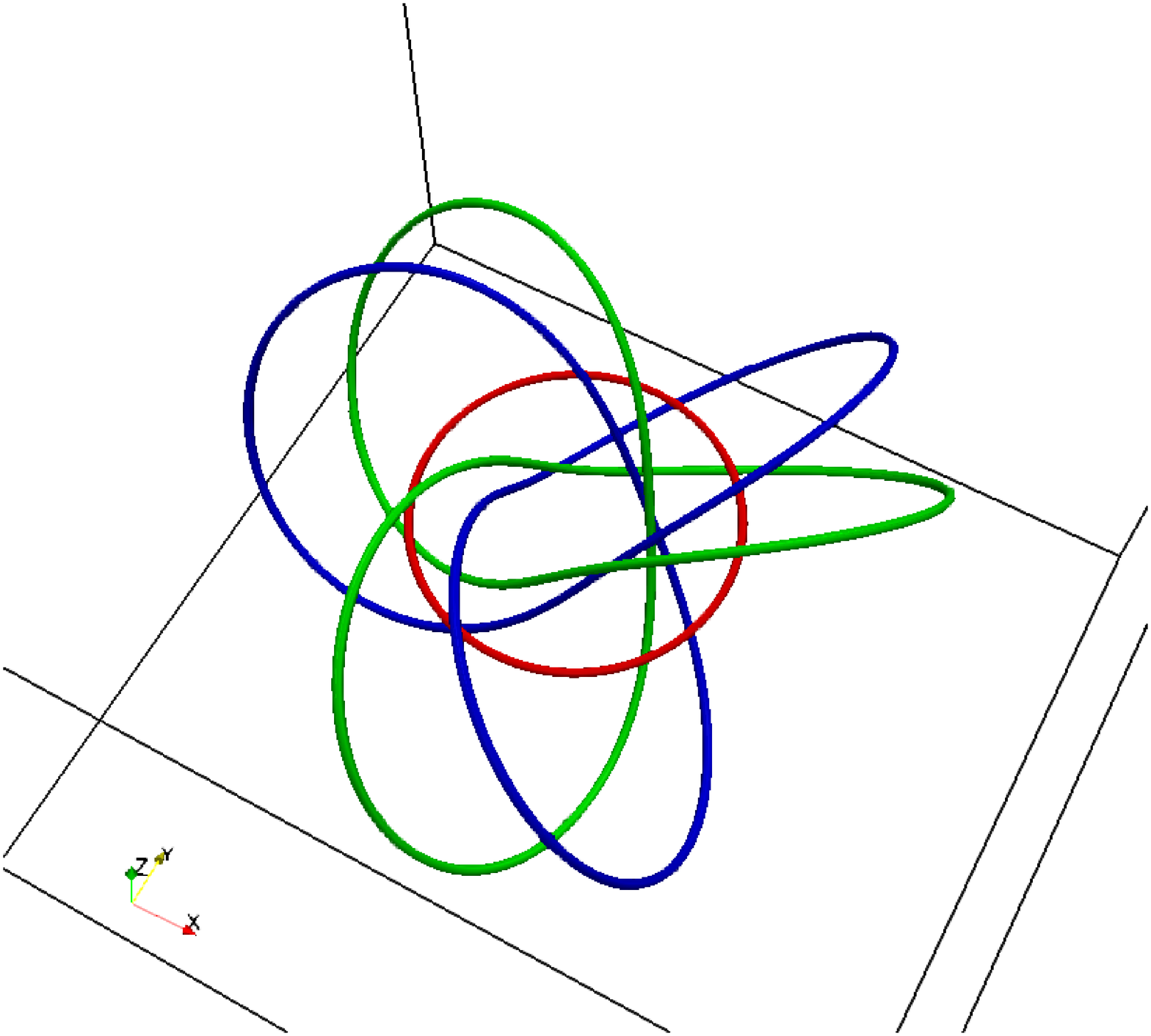}
\end{center}
\caption[]{
Several field lines of the initial magnetic field for the Hopf field with $\omega_1 = \omega_2 = 1$
(upper panel) and the field with parameters $\omega_1 = 3$, $\omega_2 = 2$ (lower panel).
We show select field lines with the ring at $||\xx|| = 1, z = 0$ (red) and
two more field lines (blue and green).
The upper field consists of linked magnetic flux rings, while the lower
consists of linked trefoil knots.
}
\label{fig: B0}\end{figure}

The vector field of this localized, finite-energy magnetic field with winding numbers
$\omega_1$ and $\omega_2$ is given by:
\EQ
\BB_{\omega_1, \omega_2}=\frac{4\sqrt{s}}{\pi (1+r^2)^3\sqrt{\omega_1^2+\omega_2^2}}
\begin{pmatrix} 2(\omega_2  y- \omega_1 xz  ) \\ -2( \omega_2  x + \omega_1 yz) \\ \omega_1(-1+x^2+y^2-z^2) \end{pmatrix},
\label{eq: hopf field}\EN
with $r^2 = x^2 + y^2 + z^2$ and $s$ is a scaling factor.
The derivation of equation \eqref{eq: hopf field} is given in appendix \ref{sec:appendixHopf}.
Selected field lines for the $\BB_{1,1}$ and $\BB_{3,2}$ fields are shown in \Fig{fig: B0}.

In recent resistive numerical simulations \cite{Smiet-Candelaresi-2015-115-5-PRL} it was shown
that magnetic fields consisting of initially linked field lines relax to an equilibrium where
field lines lie on nested toroidal surfaces.
The rotational transform (or $q$-factor) was seen to be within 10\% constant for every magnetic surface
in the structure.
The field generated by equation \eqref{eq: hopf field} consists of field lines on nested
toroidal surfaces with a constant rotational transform, which is determined by
$\omega_1/\omega_2$, and thus is topologically similar to the fields observed in resistive
simulation.
Even though the fields have similar magnetic topology, the geometrical distribution of field is
different.
The magnetic field given by equation \eqref{eq: hopf field} is not in equilibrium, as there are
large rotational Lorentz forces that cannot be balanced.
Using a \refbold{topology-conserving} relaxation scheme we will see how these forces relax the magnetic field to a
different geometry, but with the exact same magnetic topology.

\begin{figure}[t!]\begin{center}
\includegraphics[width=0.9\columnwidth]{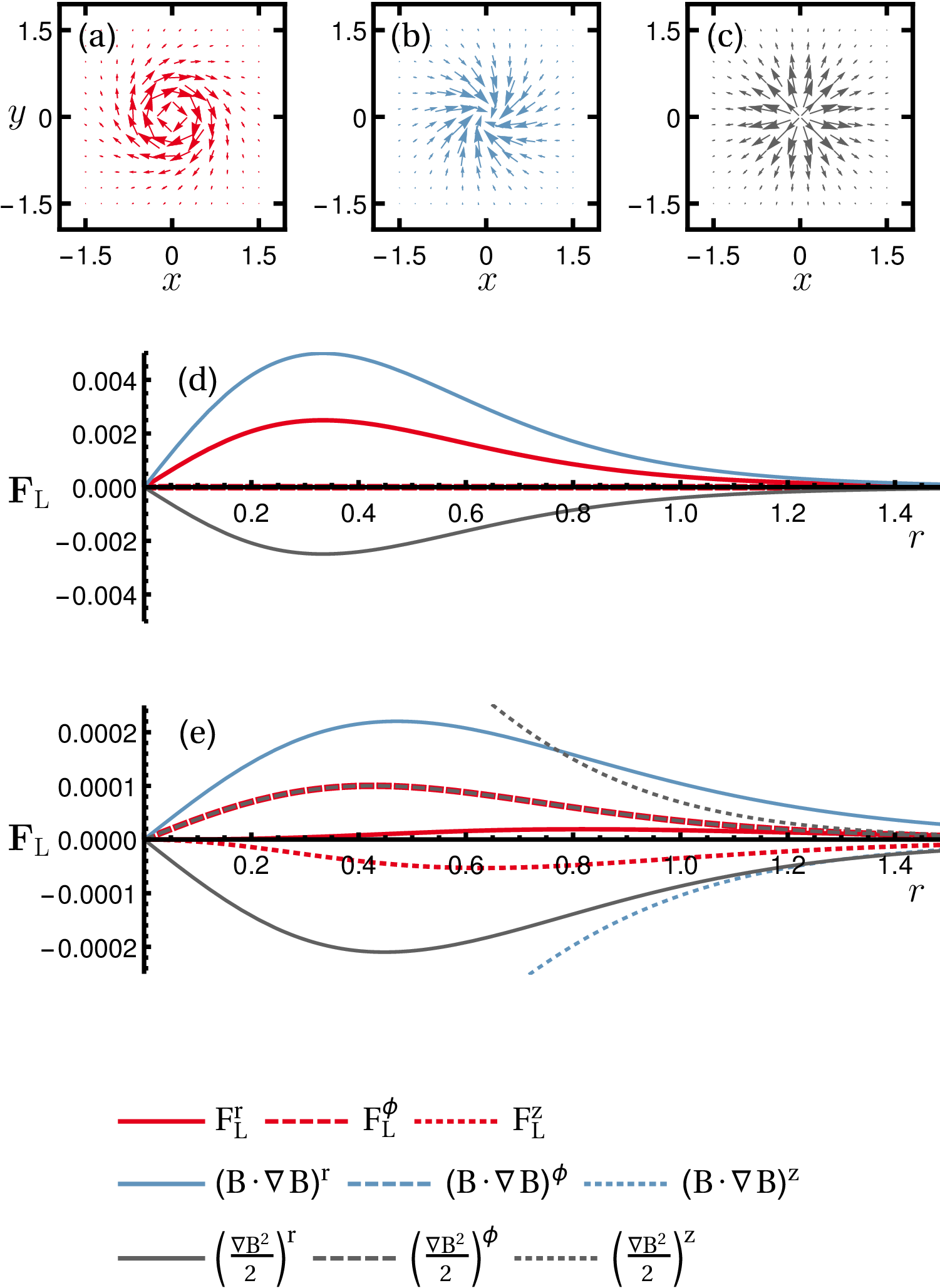}
\end{center}
\caption[]{
The Lorentz force and its components of magnetic pressure and magnetic tension for the Hopf field.
(a-c) vector plots of the Lorentz force (red), magnetic tension (blue), and magnetic pressure force
(grey) in the plane $z=1$.
(d) Lorentz force and its components in the $z=0$ plane, where there is only a radial component.
(e) Lorentz force and its components in the $z=-1$ plane.
The $\phi$ and $\theta$ components of the magnetic tension force and the magnetic pressure force
cancel each other to a large degree, leaving mainly the $\phi$ component.
Because of the symmetry of the Hopf field, the components not shown in (a-c) can be read from (e).
}
\label{fig: forces}
\end{figure}

The Lorentz force $\FF_{\rm L} = \JJ \times \BB$ can be decomposed as
$\FF_{\rm L} = \BB \cdot \nabla \BB - \nabla \BB^2/2$, where  $\BB^2/2$ is
referred to as magnetic pressure, and $\BB \cdot \nabla \BB$ is called magnetic tension.
Magnetic pressure gives rise to a force pointing
from regions with high magnetic field energy to regions of low magnetic energy.
In the Hopf fibration  magnetic energy is highly localized ($B_{1,1}^2=16/(\pi^2 (1 + r^2)^4)$),  giving rise to
a radial outward force.
The magnetic tension force, on the other hand, is a force that resists the bending of
magnetic field lines, and can effectively be seen as the result of tension in
the field lines.
\Fig{fig: forces} shows how the magnetic tension and pressure interact
to produce the Lorentz force in the Hopf field $\BB_{1,1}$.
In the $z = 1$ plane the tension adds a clockwise twist to the field and a force
toward the center, whereas the magnetic pressure
points radially outward.
The radial components largely cancel resulting in a predominantly rotational force around the
$z$-axis.
In the $z =0$ plane the forces only have a radial component, resulting in a net outwards force,
and in the $z = -1$ plane the toroidal forces are opposite with respect to $z = 1$.

\subsection{Relation to the Kamchatnov-Hopf Soliton}
The magnetic field in equation \eqref{eq: hopf field} was used by Kamchatnov
\citep{kamchatnov1982topological} to describe an ideal MHD soliton,
a solution to the ideal, incompressible MHD equations.
By setting the fluid velocity equal to the (local) Alfv\'en speed,
\EQ
\uu_{\rm Kam}=\pm\frac{\BB}{\sqrt{\rho}}
\EN
(a solution shown by Chandrasekhar
to be stable \citep{chandrasekhar1956stability, chandrasekhar1961hydrodynamic}) and using the pressure
\EQ
p_{\rm Kam}=p_\infty - \frac{B^2}{2},
\EN
it follows from the ideal induction equation
\EQ \label{eq: ideal induction}
\frac{\partial \BB}{\partial t} = \nab\times(\uu\times\BB),
\EN
that the magnetic field is static.
If we write the momentum equation as
\EQ \label{eq:motionrewrite}
 \frac{\partial \uu}{\partial t} + \uu\cdot\nabla\uu - \frac{1}{\rho}\BB\cdot\nabla \BB
 + \frac{1}{\rho}\nabla \left(p + \frac{B^2}{2} \right) = 0,
\EN
and fill in the value for $\uu_{\rm Kam}$ and $p_{\rm Kam}$ with $\BB_{1,1}$ as a magnetic field this reduces to $\partial \uu/\partial t = 0$, a static configuration.

Kamchatnov's construction solves the ideal, incompressible MHD equations,
but this solution requires a fluid velocity parallel to the magnetic field at every point
in space.
Furthermore, it is necessary in Kamchatnov's construction to include a confining pressure
$p_\infty$.  From the virial theorem we know that an external pressure can provide a restoring
force so that a simpler equilibrium, without parallel fluid flow can be achieved.  Furthermore,
we need not restrict ourselves to the case of incompressible MHD, but we look for an
equilibrium in the more general case of compressible \refbold{barotropic} ideal MHD.
In our work we will consider the \refbold{topology preserving, compressible relaxation} of the
magnetic
field starting with the Hopf map.
The field will relax to a different geometry but the topology preserving evolution will
guarantee that the field
remains topologically identical to the Hopf fibration.

\section{Methods}
In order to simulate the topology conserving relaxation we restrict the field's
evolution to such that follow the ideal
induction equation given in equation \eqref{eq: ideal induction}.

For the velocity field we use, depending on the case, two different
approaches.
In the magneto-frictional \cite{Chodura19813d} approach the velocity is
proportional to the forces on the fluid element:
\EQ \label{eq: magneto frictional}
\uu = \JJ\times\BB - \cs^2\nab \rho,
\EN
with the electric current density $\JJ = \nab\times\BB$
and sound speed $\cs$.
The sound speed effectively determines the pressure in the simulation through $p = \cs^2\rho$.
It was shown by \cite{Craig-Sneyd-1986-311-451-ApJ} that the magneto-frictional approach reduces
the magnetic energy strictly \refbold{monotonically}.
Alternatively, we can use an inertial evolution equation for the velocity
\citep{relaxation2015} with
\EQ \label{eq: inertia}
\frac{\dd\uu}{\dd t} = (\JJ\times\BB - \cs^2\nab\rho - \nu\uu)/\rho,
\EN
with the damping parameter $\nu$.

Numerical methods using fixed grids and finite differences typically introduce
numerical dissipation which would effectively add the term $-\eta_{\rm num}\JJ$ on the
right hand side of equation \eqref{eq: ideal induction}, with the numerical resistivity
$\eta_{\rm num}$ over which there is little to no control.
For every finite value of $\eta_{\rm num}$, however small, the field will invariably
undergo a change in topology.
To circumvent this we make use of a Lagrangian grid where the grid points move with the
fluid \citep{Craig-Sneyd-1986-311-451-ApJ, Candelaresi-2014-mimetic}
\EQ \label{eq: grid evolution}
\frac{\partial \yy(\xx, t)}{\partial \xx} = \uu(\yy(\xx, t), t),
\EN
with the initial grid positions $\xx$ and positions at later times $\yy$.
\refbold{The magnetic field on the distorted grid can be computed as the pull-back of a
differential $2$-form, which then leads to the simple form
(see for example references \citep{Craig-Sneyd-1986-311-451-ApJ, Candelaresi-2014-mimetic}):}
\EQ \label{eq: magnetic field pull-back}
B_{i}(\xx, t) = \frac{1}{\Delta}\sum_{j = 1}^{3} \frac{\partial y_{i}}{\partial x_{j}}
B_{j}(\xx, 0),
\EN
with $\Delta = \det{\left(\frac{\partial y_{i}}{\partial x_{j}}\right)}$.

We choose line tied boundary conditions where the velocity is set to
zero and the normal component of the magnetic field is fixed.
To compute the curl of the magnetic field $\JJ = \nab\times\BB$ on the
distorted grid we make use of mimetic spatial derivatives which increases
accuracy and ensures $\nab\cdot\nab\times\BB = 0$ up to machine precision
\cite{hyman1997natural,hyman1999mimetic}.

It should be noted that equations \eqref{eq: magneto frictional} and \eqref{eq: inertia} are
both different from the momentum equation \eqref{eq:motionrewrite},
and that therefore the evolution of the field
\refbold{is different from the evolution of a system adhering to the
(dissipationless) ideal MHD equations}.
Nevertheless, it is clear that when the relaxation reaches a steady state, either by
equation \eqref{eq: magneto frictional} or by \eqref{eq: inertia},
the field has reached a configuration in which all forces cancel.
Our evolution equation also does not conserve energy, as any fluid motion is damped in order to
expedite convergence to equilibrium.
Since our main interest is investigating the existence and character of the equilibrium that is achieved
under conservation of field line topology, the magneto-frictional and inertial evolution are
both valid approaches to achieve this equilibrium.
\refbold{A different method which uses a Hamiltonian formulation for the field, and allows for
relaxation under conservation of additional invariants, albeit under reduced dimensionality is
found in \cite{chikasue2015simulated}.}

Equations \eqref{eq: magneto frictional}, \eqref{eq: inertia}, \eqref{eq: grid evolution}
and \eqref{eq: magnetic field pull-back} are solved with the
numerical code {\textsc GLEMuR} \citep{Candelaresi-2014-mimetic, glemur-github},
written in CUDA and which runs on graphical processing units.

\section{Topology Preserving Relaxation}
We perform numerical experiments with the Hopf field as initial condition
(eq.\ \eqref{eq: hopf field}) for different parameters $\omega_1$ and $\omega_2$,
and the scaling factor $s$.
The initial density is constant in space resulting in a constant pressure set by $\cs^2$.
All the simulations conserve the topology and obey either the magneto-frictional
equation of motion \eqref{eq: magneto frictional} or the momentum
equation \eqref{eq: inertia}.

\subsection{Field Expansion}
We first analyze the relaxation of the $\BB_{1,1}$ field with $s = 2$.
As can be expected from the distribution of forces in the initial field (\Fig{fig: forces})
the field expands outwards in the $xy$-plane, whilst the grid is twisted in opposite
directions in the $z=1$ and $z=-1$ planes.
The motion of the grid for the magneto-frictional runs with $\cs^2=0.1$  are shown in
supplemental videos 1 and 2 and in \Fig{fig: grid distortion} (Multimedia view).
Supplemental video 1 shows the displacement of the grid initially in the $z=1$ plane,
which twists in a clockwise direction.
The colors indicate the vertical displacement of the grid, which moves towards the origin
in the center, and upwards further out.
If we look at the motion of the grid in the $y=0$ plane (supplemental video 2), we see the
grid expanding outwards in the $z=0$ plane.
The grid spacing increases around the $z=0, \ x=1$ location, resulting in the
formation of a region of lowered pressure.
As the field lines move with the grid, this is also the new location of the degenerate
field line.

\begin{figure}[t!]\begin{center}
\includegraphics[width=0.9\columnwidth]{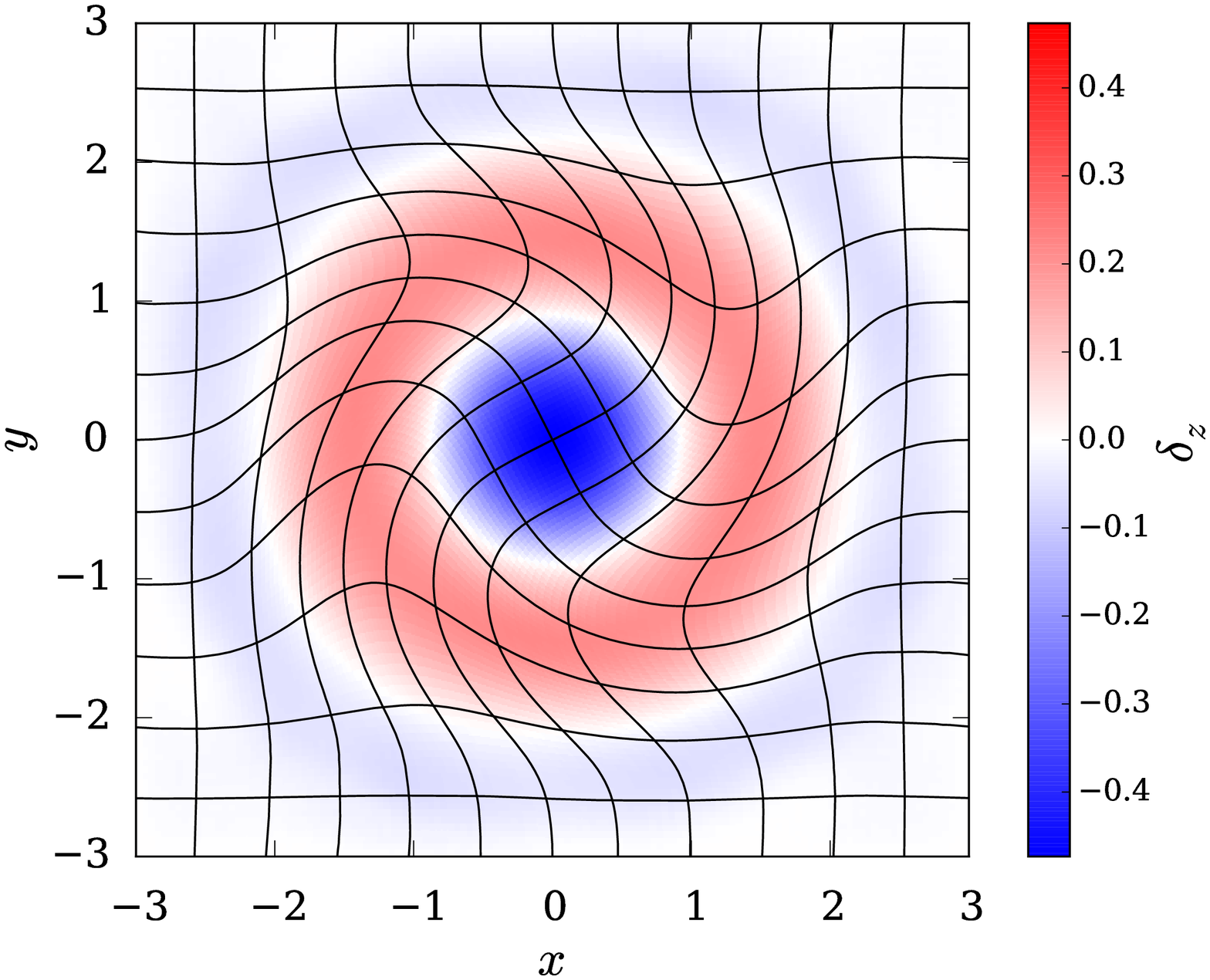} \\
\includegraphics[width=0.9\columnwidth]{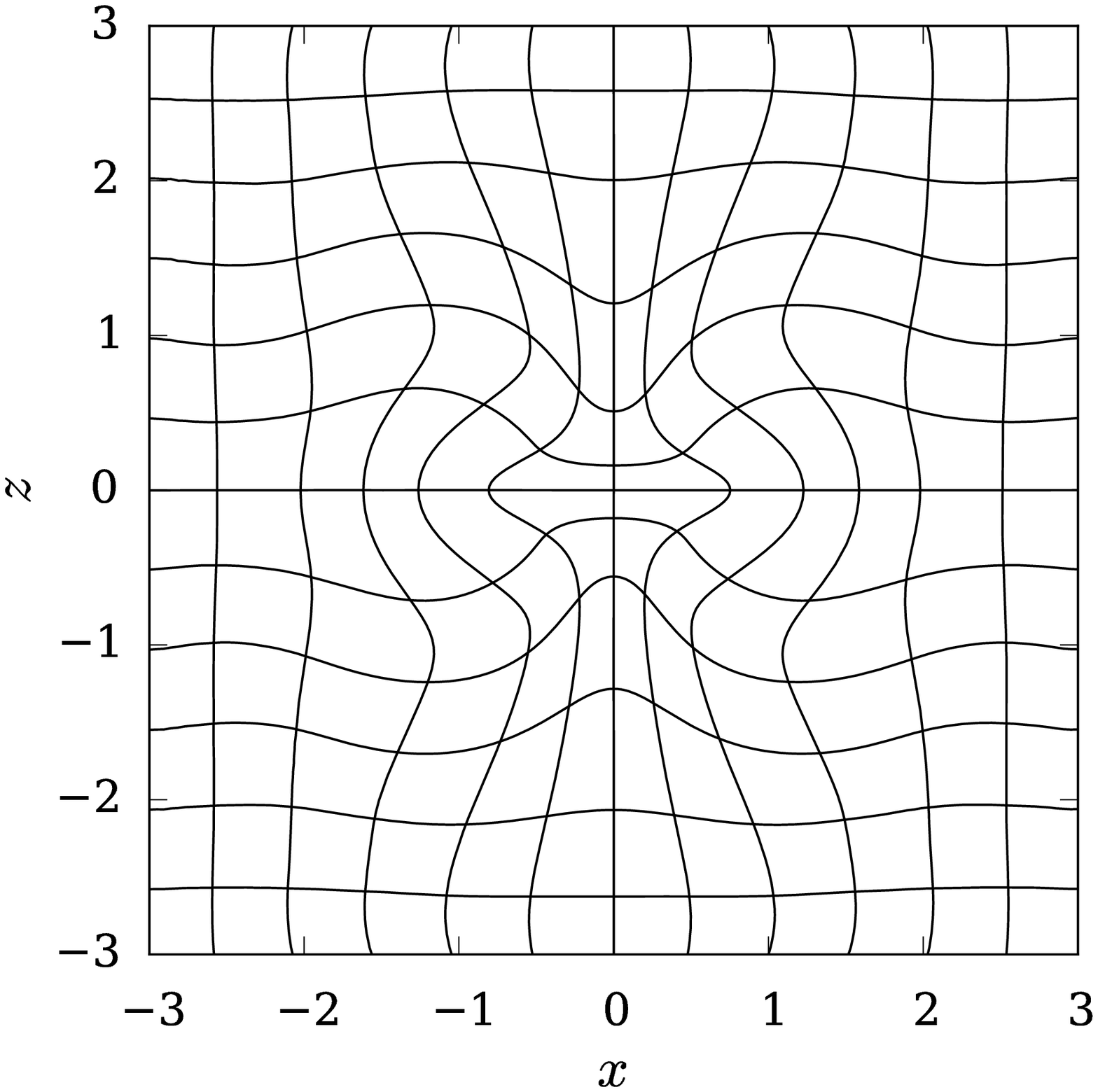}
\end{center}
\caption[]{
    Grid distortion during relaxation to the final, relaxed configuration (approx. time $t = 200$). (Upper panel): points initially in the  $z = 1$ plane
(multimedia view), and (lower panel): distortion of the the $y = 0$ plane (multimedia view).
The color denotes the deviation of the grid points in the $z$-direction compared to $t = 0$.
}
\label{fig: grid distortion}\end{figure}

\begin{figure}[t!]\begin{center}
\includegraphics[width=0.9\columnwidth]{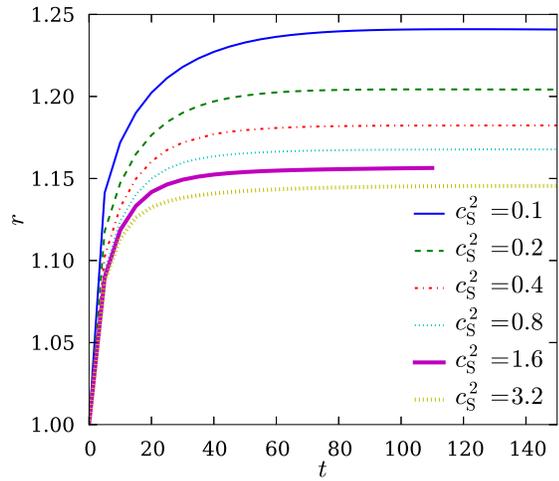} \\
\includegraphics[width=0.9\columnwidth]{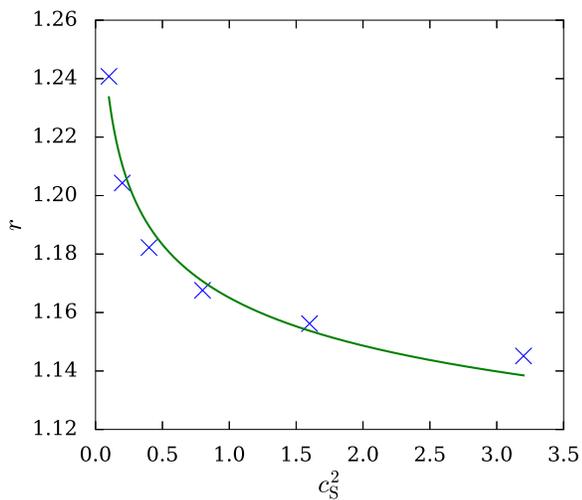}
\end{center}
\caption[]{
Time evolution of the degenerate field line in $\BB_{1,1}$ at different effective pressures using
the magneto-frictional approach (upper panel).
Radii at time $t = 150$ for different values of $\cs^2$ with fit (lower panel).
}
\label{fig: radius expansion}\end{figure}

The expansion in the $xy$-plane can be tracked by measuring the change in radius $r$ of
the degenerate field line.
This is measured by the displacement of the point initially at $(1,0,0)$, and shown in
\Fig{fig: radius expansion} (upper panel) for several different effective pressures.
The effective pressure is set by the parameter $\cs^2$, which enters into the equations as the proportionality factor between
density $\rho$ and pressure $p$.
For values lower than $\cs^2=0.1$ the field expands to the computational boundaries.
For higher values of $\cs^2$ we see that, as expected, the expansion of the field levels off after a certain
time, and the higher the confining pressure is, the less the configuration expands
before it reaches equilibrium.

For $\cs^2 = 0$ we expect an unconstrained expansion, while in the limit of
$\cs^2 \to \infty$ we should see no expansion.
Therefore, we plot the radius $r$ vs.\ $\cs^2$ at time $t = 100$
and fit the function
\EQ
r = b(\cs^2)^{a} + 1,
\EN
with fitting parameter $a = -0.160494$ and $b = 0.16229561$.
This fit gives a reasonable approximation for the expansion of the degenerate field line,
indicating how the radius of the relaxed configuration depends on confining pressure.
(\Fig{fig: radius expansion}, lower panel).

\begin{figure}[t!]\begin{center}
\includegraphics[width=0.9\columnwidth]{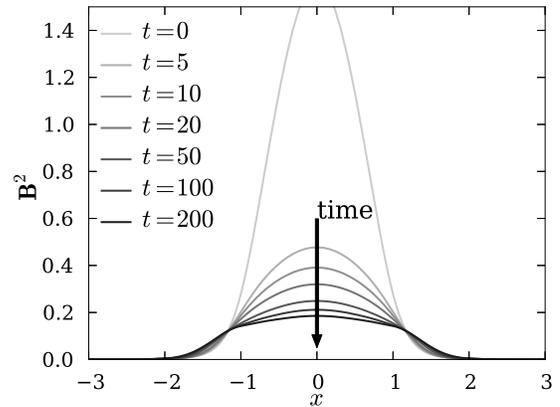}
\end{center}
\caption[]{
Squared of the magnetic field strength $B^2$ on the $x$-axis for different times for the  $\BB_{1, 1}$ field.
The field was relaxed using the magneto-frictional approach with $\cs^2 = 0.2$.
The magnetic field strength, and hence the magnetic pressure force, is greatly reduced during
the relaxation by plasma expansion perpendicular to the field direction.
}
\label{fig:_B2}\end{figure}

During this expansion, the magnetic energy $B^2$ in the configuration sharply decreases
due to the plasma expansion perpendicular to the magnetic field direction.
This process can be seen in \Fig{fig:_B2}, and it causes a drastic decrease in the
magnetic pressure from the initial configuration.

\subsection{Force Balance}
\begin{figure}[t!]\begin{center}
\includegraphics[width=0.9\columnwidth]{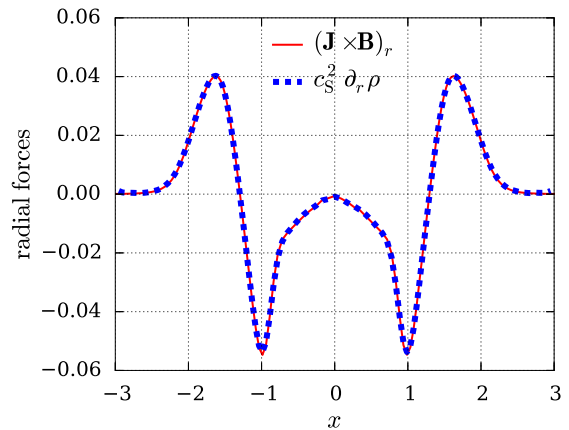}
\end{center}
\caption[]{
Radial component of the Lorentz force and radial component of the pressure gradient along the $x$-axis.
The field was relaxed using the inertial approach with $\cs^2=0.1$ and $\nu=1$.
The two forces balance each other almost perfectly, indicating that an equilibrium is reached.
}
\label{fig:_forces}\end{figure}

Our simulations relax to a static configuration where the fluid velocity is zero.
From the momentum equation \eqref{eq:motionrewrite}, we can see that for any
static equilibrium the pressure forces have to be balanced by a gradient in pressure:
\EQ \label{eq: force balance}
\JJ\times\BB = \nab p (= \cs^2 \nab \rho).
\EN
If we look at the relaxed field we see that the pressure is no longer constant,
but the plasma has reorganized to create a toroidally-shaped region of lower pressure.
The Lorentz force is also different in the relaxed configuration.
The magnetic pressure contribution has been greatly reduced by the lowering of magnetic
field strength accompanying the expansion, and the Lorentz force is now directed outwards, away
from the degenerate field line.
The condition of force balance in equation \eqref{eq: force balance} is achieved in
 the simulation run, as can be seen in \Fig{fig:_forces}.
The Lorentz force $\JJ \times \BB$ is balanced by the pressure force
$-\nabla p= -\cs^2 \nabla \rho$, such that the total force is zero.

\begin{figure}[t!]\begin{center}
\includegraphics[width=0.9\columnwidth]{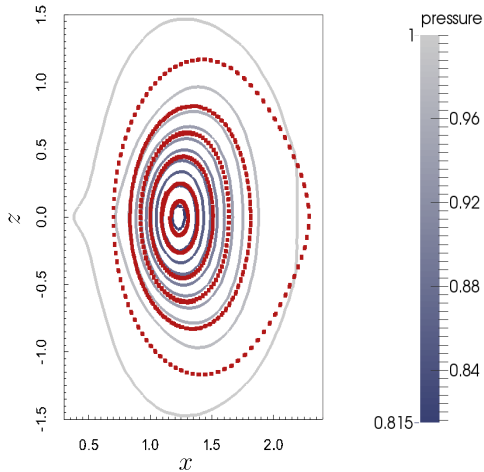}
\end{center}
\caption[]{
Magnetic surfaces (red squares) and pressure contours (colored lines) in the $xz$-plane for for the
relaxed $\BB_{1,1.01}$ field.
The inner magnetic surfaces coincide with the pressure surfaces.
Because the pressure gradients and Lorentz force are much lower on the outer surfaces,
convergence to the equilibrium state is much slower.
}
\label{fig: shafranov}\end{figure}

Another consequence of the equilibrium condition is that the pressure
must be constant on magnetic field lines, and thus on the toroidal surfaces
on which the field lines lie.
By construction every field line in the Hopf field is a closed circle, but the circles
lie on the surfaces of nested tori.
These surfaces become visible if we consider the field with parameters $\omega_1=1$ and
$\omega_2=1.01$, such that every field line is a $(100,101)$ torus knot.
This field is locally nearly indistinguishable from the $\BB_{1,1}$ field,
but by tracing a single field line
the toroidal surface on which the field line lies becomes visible.
By plotting the intersections of such a field line with the $xz$-plane (constructing a Poincar\'e plot), we see a cross section
of the magnetic surfaces.
These magnetic surfaces are plotted together with the contours of constant pressure in the relaxed
magnetic configuration in \Fig{fig: shafranov}.
The contours of constant pressure clearly conform to the shape of the magnetic surfaces,
especially near the degenerate field line.
We attribute the discrepancy between the outermost magnetic surfaces and pressure surfaces
to the fact that both the pressure gradient and/or the magnetic field strength are low at these locations, leading to a
slow magneto-frictional convergence to equilibrium.

Since the initial configuration is axisymmetric, and the resultant forces are as well,
the configuration will remain axisymmetric through the entire evolution.
The topology preserving relaxation method thus relaxes the magnetic field to an
axisymmetric configuration where the magnetic forces are balanced
by the pressure gradient.
This kind of equilibrium can, in principle, be described by a solution to
the Grad-Shafranov equation\citep{shafranov1966plasma}, but finding the exact functional form
is a non-trivial task.

From the combination of the lowering of magnetic energy shown
in \Fig{fig:_B2} and the lowering of the pressure in a toroidal region
as seen in \Fig{fig: shafranov} we understand how the equilibrium is
achieved in light of the virial theorem.
Recall from equation \eqref{eq: virial tensor} that the volume contribution consists of
three terms, $\rho u^2$, $3p$, and $B^2$.
The contribution of the velocity is zero in equilibrium.
As the field expands, the relative contribution of
$B^2$ drops.
The fluid that is expelled from the toroidal region causes a slight increase in pressure
distributed over the entire surface.
An equilibrium can be achieved when this negative contribution balances the reduced magnetic
pressure of the distorted field.

The force-balanced equilibrium state obtained in these
ideal relaxation experiments bears strong resemblance with quasi-stable
magnetic structures found in various recent simulations, such as
magnetic bubbles in\citep{gruzinov2010solitary}, freely decaying relativistic
turbulence in\citep{zrake2016freely}, and self-organizing knotted magnetic structures
in\citep{Smiet-Candelaresi-2015-115-5-PRL}.

\subsection{Dependence on $\omega_1$ and $\omega_2$}
To investigate the effects of different field line topologies we simulate
the ideal relaxation of $\BB_{3,2}$ and $\BB_{2,3}$ with $\cs^2 = 0.1$ and the scaling
factor $s=1$.
These two fields have exactly the same magnetic energy, but their magnetic
topology, and the spatial distribution of magnetic pressure is different.
In $\BB_{3,2}$ the field lines make 3 poloidal (short way around the torus) windings
for two toroidal windings.
If we look at equation \eqref{eq: hopf field}, we can see that $\omega_1$ (responsible for the poloidal winding)
multiplies the $z$-component of the field, and increases the field strength along the $z$-axis of
the configuration, whereas $\omega_2$ increases the magnetic pressure around the degenerate torus
in the $xy$-plane.

\begin{figure}[t!]\begin{center}
\includegraphics[width=0.9\columnwidth]{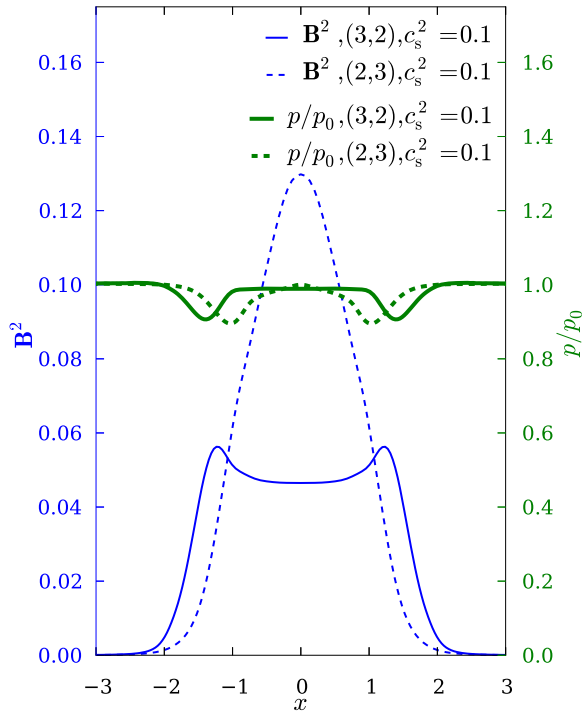}
\end{center}
\caption[]{
Magnetic energy density and normalized pressure
on the $x$-axis for simulation runs with $\cs^2 = 0.1$
and different ratio of poloidal to toroidal winding.
The magnetic energy distribution is different in the two
relaxed configurations, with the $\BB_{3,2}$ simulation showing
highest magnetic field strength around the degenerate torus, and
the $\BB_{2,3}$ configuration the highest field strength around the
$z$-axis.
}
\label{fig: b2_tf_compare_32}\end{figure}

When we relax the field we see that both choices of $\omega_1$ and $\omega_2$ yield
an equilibrium, but the magnetic energy and pressure
distributions are different, as can be seen in \Fig{fig: b2_tf_compare_32}.
The radial expansion of the $\BB_{3,2}$ simulation is much larger than that of $\BB_{2,3}$,
indicating that the degenerate torus (located at the minimum in pressure) is pushed further outwards.
Note that the $\BB_{3,2}$ equilibrium, which started out with relatively higher magnetic
pressure on the $z$-axis,
now shows highest field around the degenerate torus.
The $\BB_{2,3}$ field now has a highest magnetic field strength around the origin.

We can intuitively understand the behavior of these fields by recalling a well known
observation in MHD; under internal forces a magnetic flux ring contracts and fattens,
whilst a ring of current becomes thinner and stretches\citep{bellan2008fundamentals}.
A ring of current gives rise to a magnetic field with only poloidal magnetic field lines, whereas a ring of magnetic flux
consists of purely toroidal magnetic field lines.
The fields we consider lie in between these two extreme configurations.
The stronger the poloidal winding, the more the configuration resembles a current ring, and therefore this configuration
will stretch relatively more.
This will leave a relatively high magnetic field around the degenerate torus, as we can see in the
equilibrium achieved by the $\BB_{3,2}$ field.
The $\BB_{2,3}$ field has a relatively higher toroidal field, and will therefore expand less, leaving
a high field around the $z$-axis.

Even though the exact distribution of magnetic energy and the magnetic field topology are different, the
equilibrium is always characterized by a toroidal region of lowered pressure.
All simulations start with constant hydrostatic pressure, and the observed final magnetic energy
distribution is then given by the deformation that balances the dip in hydrostatic pressure with the
lowered magnetic pressure.
A different initial pressure distribution would also result in qualitatively different equilibrium  magnetic
energy distributions, but the essential features of the equilibrium would remain the same.

\subsection{Force-Balance with a mean magnetic field.}
As noted in the introduction, it is possible to balance the field if
the contribution of $\mathbf{T}_{\BB}$ is non-zero at the boundary, i.e.\ the field
is balanced by a finite external magnetic pressure.
We investigate this by evolving $\BB_{1,1}$ in a weak background field
$\BB_{\rm bck} = -0.02\ee_z$ such that the final field is $\BB=\BB_{1,1}+\BB_{\rm bck}$.
It should be noted that this background field changes the magnetic topology
of the initial condition.
The new magnetic topology is such that field lines far away from the $z$-axis, where
the field strength is opposite, but weaker than the guide field do not form magnetic surfaces, but extend from $z=-\infty$ to $z=\infty$.
The same is the case for field lines close to the $z$-axis.
On the magnetic surfaces that remain toroidal the ratio of poloidal to toroidal winding
now changes from surface to surface.

\begin{figure}[t!]\begin{center}
\includegraphics[width=0.9\columnwidth]{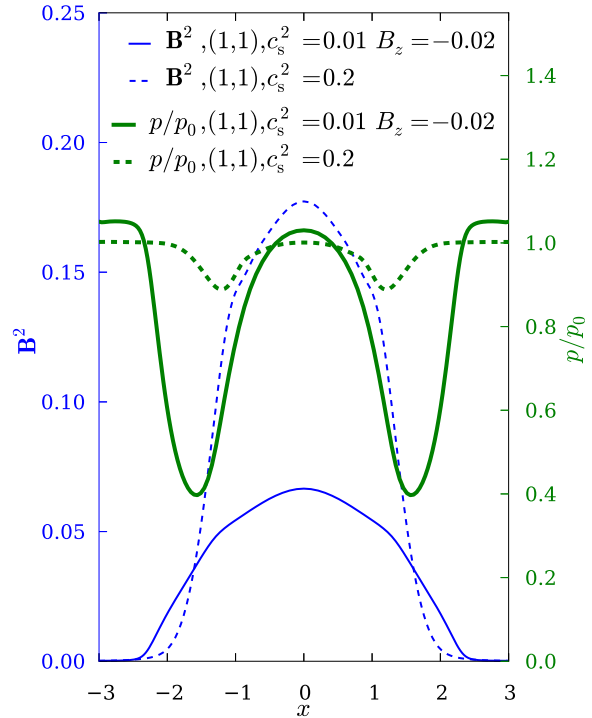}
\end{center}
\caption[]{
Magnetic energy density and normalized pressure
on the $x$-axis for simulation runs with $1:1$ ratio of poloidal to toroidal winding
with and without background magnetic field.
}
\label{fig: b2_tf_compare_11}\end{figure}

For our numerical experiment we reduce the effective pressure by setting $\cs^2 = 0.01$.
This setting is much too low for the magnetic field to reach equilibrium
within the simulation box without a guide field, but with the guide
field an equilibrium is reached.
The density and magnetic energy distribution are shown in \Fig{fig: b2_tf_compare_11}.

As the field expands, it pushes the guide field outwards, creating a restoring
magnetic tension force.
At the same time the magnetic field strength decreases, and the external magnetic
pressure force halts the expansion.
A finite, although very low, effective pressure is necessary to prevent the field from
expanding indefinitely in the direction of the field lines.
As we can see, the region of lowered pressure is much larger in the
guide field simulation, which is to be expected from such a low value of
$\cs^2$.
This numerical result suggests that in three dimensions a localized magnetic excitation
can only achieve equilibrium if there is a finite external pressure.
We note that this is in contrast to promising results found in two dimensional
simulations \citep{zrake2016freely,gruzinov2010solitary}, where localized magnetic excitations,
called magnetic bubbles, are found in zero-pressure MHD and Force-Free Electrodynamics.

\section{Discussion and conclusions}
We have shown how a localized magnetic excitation, in particular a Hopf field, relaxes
\refbold{to a configuration which is an equilibrium in an ideal, compressible plasma.}
\refbold{The virial theorem implies} that for any equilibrium to exist,
there must either be a finite external pressure, or a guide field to attain equilibrium.
The equilibrium that is achieved consists of a toroidal depression and is not a Taylor
state.

We have used a topology preserving Lagrangian relaxation scheme that converges
to an equilibrium configuration and observe the equilibrium in a wide range of
parameters and different realizations of the Hopf map.
In contrast to the topological solitons described by Kamchatnov\citep{kamchatnov1982topological},
these configurations are static,
and do not require a fluid velocity to balance the equations.
These configurations  are therefore static topological \refbold{solitons} in compressible MHD.
The magnetic configurations remain axisymmetric under time evolution, and
an equilibrium is achieved when magnetic field lines conform to the toroidal surfaces of
constant pressure.
The Lorentz force is balanced by the gradient in pressure and the
obtained equilibria can be considered
Grad Shafranov equilibria \citep{shafranov1966plasma}.
Changing the magnetic topology of the initial field, by adjusting the ratio of toroidal to
poloidal winding
yields a qualitatively similar equilibrium, with a different distribution of magnetic energy.

Recent numerical simulations have shown that localized helical magnetic configurations
can be generated in resistive plasma\citep{Smiet-Candelaresi-2015-115-5-PRL, gruzinov2010solitary}.
The equilibrium we observe here is similar to what is observed in the resistive simulations,
except that the ideal relaxation conserves field line topology, and therefore
magnetic islands cannot be created.

Even though an equilibrium at zero pressure is impossible,
any realistic plasma in which a topologically nontrivial
field is embedded will have a (possibly very low) finite pressure.
If such a field exists in a close to ideal plasma, the expansion will
cause a decrease in magnetic field magnitude, and corresponding magnetic pressure.
A finite external pressure, no matter how low, will give rise to an
equilibrium at which the external magnetic pressure is able to
confine the magnetic field in the manner described in this paper.
Examples of where this could occur are in experiments with plasmoids
\cite{bostick1956experimental, armstrong1980compact,  perkins1988deep, wright1990field},
and in astrophysical plasma such as the magnetic bubbles studied by
Braithwaite\cite{braithwaite2010magnetohydrodynamic}.

It is interesting to contrast the equilibrium found in our simulations to the localized
magnetic bubbles described in\citep{gruzinov2010solitary, zrake2016freely}.
The authors found a localized increase in pressure in two-dimensional relaxation at zero pressure, but
as we have shown, such equilibria are impossible in three dimensions due to expansion
along the guide field.

Localized three-dimensional magnetic excitations are possible, and the tell-tale signature
of these relaxed states is a toroidal lowering of plasma pressure coinciding with the innermost
magnetic surfaces.
Such signatures could be detected in astrophysical observations, and
help understanding the stability of magnetic fields in fusion plasmas.

\begin{acknowledgments}
SC acknowledges financial support from the UK's
STFC (grant number ST/K000993).
This work was supported by NWO VICI 680-47-604 and the NWO graduate programme.
The authors acknowledge support from the Edinburgh Mathematical Societies research support fund.
We gratefully acknowledge the support of NVIDIA Corporation with the
donation of one Tesla K40 GPU used for this research.
We would like to thank Gunnar Hornig, David Pontin and Alexander Russell for useful
discussions and the anonymous referees for the very useful comments.
\end{acknowledgments}

\appendix
\section{Derivation of the Hopf Field}\label{sec:appendixHopf}
If one considers the three-sphere $S^3$ embedded in $\mathbb{C}^2$ such that
$S^3=\{(z_1,z_2)|z_1\bar{z}_1+z_2\bar{z}_2=1\}$,
with $z_1, z_2 \in \mathbb{C}$
and one associates the complex plane with the sphere
$S^2$ via stereographic projection
$\pi^{(2)^{-1}}:\mathbb{C}\cup \infty \rightarrow S^2$,
then a map from $S^3$ to $S^2$ can be given by the following expression:
\EQ \label{Hopf}
h^{(\omega_1,\omega_2)}(z_1,z_2): S^3\rightarrow S^2 = \pi^{(2)^{-1}}\left({\frac{z_1^{(\omega_2)}}{z_2^{(\omega_1)}}}\right).
\EN
Here parenthesized exponentiation $z^{(\omega)}$ denotes the operation $z=r e^{i\phi} \rightarrow r e^{i\omega\theta}$ such that only the phase of the complex number is multiplied by $\omega$.
If $\omega_1$ and $\omega_2$ are equal, this map reduces to Hopf map, where every
fiber is a perfect circle and linked once with every other fiber.
This is readily checked by observing that
 $h^{(1,1)}\left( z_1,z_2 \right)= h^{(1,1)}\left( z_1 e^{i\theta},z_2e^{i\theta} \right)$,
so the fibers of the map are indeed great circles in $S^3$.
If $\omega_1$ and $\omega_2$ are not equal, but $\omega_1/\omega_2\in \mathbb{Q}$, the fibers are $(\omega_1/{\rm gcd}(\omega_1,\omega_2), \omega_2/{\rm gcd}(\omega_1, \omega_2))$ torus knots where ${\rm gcd}(a,b)$ is the greatest common divisor of $a$ and $b$.

In order to construct a field in $S^3$ from the Hopf map we modify the construction
by Ra\~nada \cite{Ranada1989}, using the method described in \cite{arrayas2014class} and
\cite{Smiet-Candelaresi-2015-115-5-PRL} by extending the Hopf map to a
complex-valued function from $S^3$ to $\mathbb{C}$:
\EQ
\phi:\mathbb{R}^3\rightarrow\mathbb{C}= \pi^{(2)} \circ h^{(\omega_1,\omega_2)}\circ  \pi^{(3)^{-1}},
\label{eq:phi}\EN
where $\pi^{(3)^{-1}}$ denotes inverse stereographic projection from $S^3$ to $\mathbb{R}^3$.

The expression for the function $\phi$ becomes:
\EQ
\phi = \frac{2(x+iy)^{(\omega_2)}}{(2z+i(r^2-1))^{(\omega_1)}},
\EN
where $r^2=x^2+y^2+z^2$. This construction is schematically illustrated
in \Fig{fig: hopfmap}.
Since, by construction, $\phi$ is constant on linked curves in $\mathbb{R}^3$, the following
expression results in a vector field that is everywhere tangent to the curves:
\EQ
\tilde{\BB}=\frac{1}{2\pi i}\frac{\nabla\phi \times \nabla\phi^*}{1+\phi\phi^*}.
\EN
This field is then given by
\EQ\label{eq:bunormalized}
\tilde{\BB} = \frac{4}{\pi(1+ r^2)^3}
\begin{pmatrix} 2(\omega_2  y- \omega_1 xz  ) \\ -2( \omega_2  x + \omega_1 yz) \\ \omega_1(-1+x^2+y^2-z^2)  \end{pmatrix}.
\EN
As a final step we normalize the magnetic field so the magnetic energy is independent of the choice
of $\omega_1$ and $\omega_2$. Since
\EQ
\int \tilde{\BB}^2 \ \mathrm{d}^3x=(\omega_1^2+\omega_2^2),
\EN
we divide equation \eqref{eq:bunormalized} by $\sqrt{(\omega_1^2+\omega_2^2)}$ to obtain
equation \eqref{eq: hopf field} in the paper, safe the scaling factor $s$.

\bibliographystyle{apsrev}
\bibliography{references}

\end{document}